\documentclass[%
 preprint, 
 amsmath,amssymb,
 aps, physrev,
]{revtex4-2}

\usepackage{graphicx}
\usepackage{dcolumn}
\usepackage{bm}
\usepackage{hyperref}
\usepackage{amsmath}

\begin{document}


\title{\textbf{Fe-Doping-Induced Band Structure Modification and Cryogenic Phase Stability in Cs$_2$AgBiBr$_6$ Single Crystals}} 

\author{Yanan Li$^{a,b}$}
\email{Corresponding author:liyanan@lntu.edu.cn}
\author{Xuejiao Wu$^b$}
\author{Jidong Deng$^{b,c}$}
\author{Jinbao Zhang$^b$}

\affiliation{$^{a}$College of Electronic and Information Engineering, Liaoning Technical University,
125105, Huludao, Liaoning, China.}
\affiliation{$^{b}$Department of Materials Science and Technology, Xiamen University, 361005, Xiamen, Fujian, China}
\affiliation{$^{c}$School of Materials and Chemistry, Southwest University of Science and Technology, 621010, Mianyang, Sichuan, China}

\thanks{This article is now published as: Yanan Li, Xuejiao Wu, Jidong Deng, Jinbao Zhang,Fe-doping-induced band structure modification and cryogenic phase stability in Cs2AgBiBr6 single crystals, Solid State Communications,Volume 409,2026,116319,ISSN 0038-1098, https://doi.org/10.1016/j.ssc.2026.116319.}




\begin{abstract}

Despite its promise as a lead-free alternative, the practical application of Cs$_2$AgBiBr$_6$ in optoelectronics is limited by its wide band gap and detrimental intrinsic defects. To overcome these challenges, we synthesized Cs$_2$AgBi$_{0.5}$Fe$_{0.5}$Br$_6$ single crystals via a modified hydrothermal method. While both pristine and Fe-doped crystals undergo a structural phase transition near 125~K, Fe incorporation fundamentally alters its impact. The dopant simultaneously narrows the band gap in the high-temperature phase and suppresses the associated cryogenic structural instability. Our optical and X-ray structural studies establish Fe doping as a powerful strategy for tailoring the properties of Cs$_2$AgBiBr$_6$, advancing its potential for high-performance, low-temperature optoelectronic and spintronic devices.

\end{abstract}

                              
\maketitle


\section{Introduction}

The search for environmentally stable and lead-free perovskites has positioned the double perovskite Cs$_2$AgBiBr$_6$ (CABB) as a promising candidate for optoelectronic applications \cite{key10,key17,key19}.  However, its practical performance is fundamentally limited by a wide energy gap, which restricts its spectral response, and the presence of intrinsic point defects, such as Bi$_{\text{Ag}}$ and Ag$_{\text{Bi}}$ antisite defects, which act as charge-trapping centers \cite{key18,key23,key24,key34,key36}. Chemical doping represents a powerful strategy to tailor these properties, yet conventional thin-film growth methods often introduce more disorder, failing to resolve these intrinsic limitations \cite{key5,key15,key37}.

A further consideration for practical device application is CABB's structural phase transition near 125 K. While the existence of this transition is recognized \cite{key23,key26}, its potential to induce significant lattice strain and degrade structural integrity during temperature cycling represents a critical, yet underexplored, challenge for device stability. Therefore, a truly advanced doping strategy must accomplish two goals: it must favorably tune the electronic band gap and enhance the crystal's resilience to phase-transition-induced strain.

In this work, we demonstrate that Iron (Fe) doping uniquely addresses both the electronic and structural challenges of CABB. We target a high doping concentration (nominally 50\% replacement of Bi) to maximize potential effects, achieved through a modified crystal growth process \cite{key15}. We show that this Fe incorporation significantly enhances the structural stability of CABB \textit{below} its 125 K phase transition temperature, suppressing its detrimental effects. Furthermore, \textit{above} the transition, the same Fe doping effectively reduces the material's energy gap. This dual-functionality is critically enabled by synthesizing high-quality single crystals, which allow for proper dopant integration without the disorder common in thin films \cite{key5}. Consequently, Fe-doped CABB single crystals emerge as a superior, robust material platform, whose inherent semiconductor character—now augmented with tailored optoelectronic properties and enhanced stability—paves the way for advanced, reliable devices in optoelectronics and spintronics.

To elucidate these effects, we synthesized high-quality single crystals of both pristine and Fe-doped CABB and systematically characterized their properties using single-crystal X-ray diffraction (SC-XRD), photoluminescence (PL), and Raman spectroscopy. Our results confirm that Fe doping maintains the fundamental crystal structure and the phase transition at $\sim$125 K. Crucially, however, we found that Fe doping serves a critical dual function. At temperatures \textit{above} the phase transition, it reduces the energy gap, as evidenced by a systematic shift in PL peak positions. \textit{Below} the transition, it markedly improves structural stability, demonstrated by a reduction in lattice strain deformation from SC-XRD. This enhanced structural quality is further corroborated by optical measurements: the Fe-doped sample exhibits a narrower and cleaner PL lineshape (reduced full-width-at-half-maximum, FWHM), while the host sample shows complex peak splitting across the transition, indicative of a more disordered and strained lattice.

\section{Results and Discussion}

\subsection{Crystal Synthesis}
To obtain high-quality single crystals, their growth was carefully optimized. While the initial synthesis followed our previous hydrothermal method \cite{key15,key10,key35}, a key modification was introduced: a five-step furnace program (Figure 1, left). This protocol was designed to optimize crystal formation through controlled heating, annealing, and slow cooling. The heating stage promoted atomic mobility and initial reaction, the annealing step facilitated homogenization and enabled the reduction of defects in the newly formed crystals, and the slow cooling prevented thermal shock \cite{key18}. Using this approach, well-defined single crystals were successfully grown, as shown in Figure 1, right. The red octahedral shape crystal corresponds to CABB, while the black one is Fe-CABB.

\begin{figure}
    \centering
        \centering
        \includegraphics[width=0.8\linewidth]{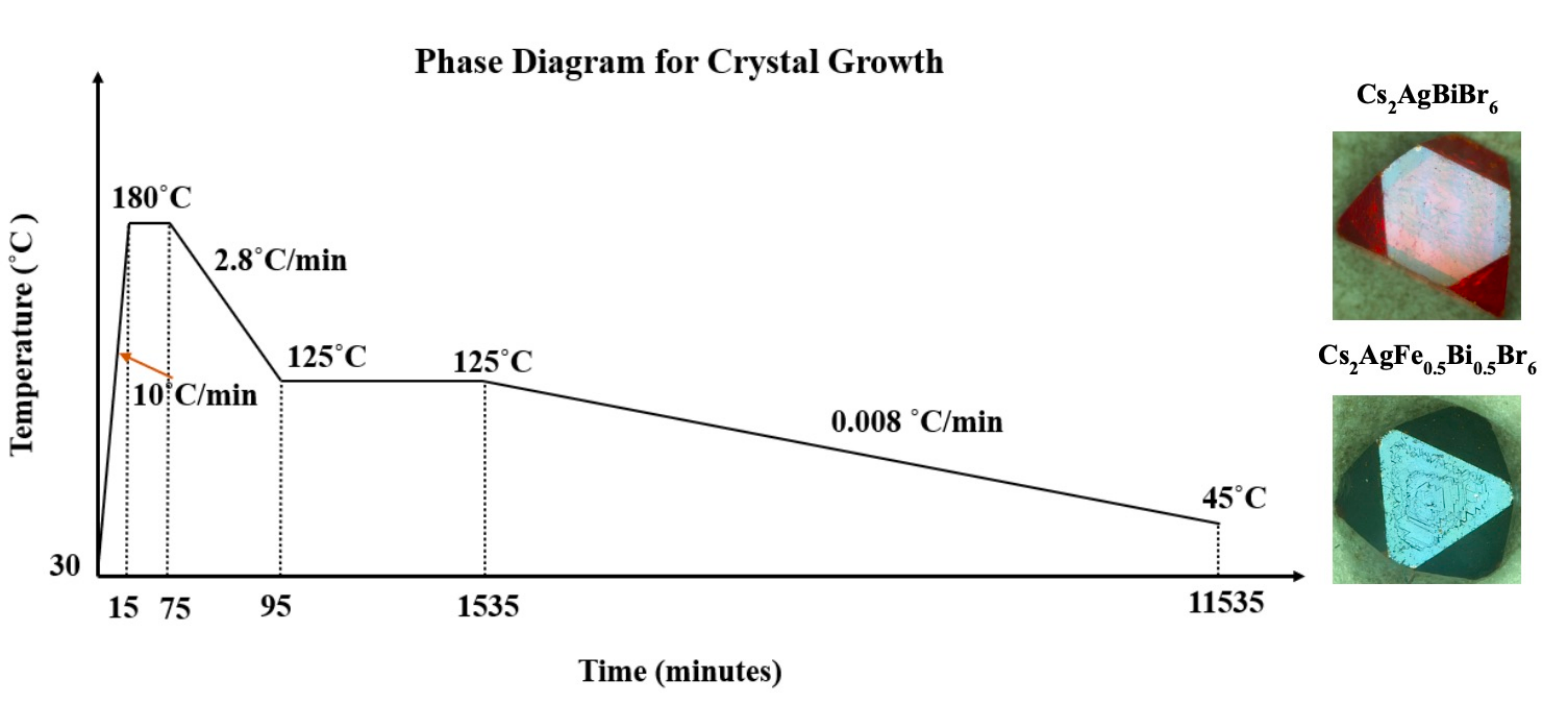}
    \caption{Phase diagram illustrating the synthesis conditions for single crystal growth of Cs$_2$AgBiBr$_6$ and Cs$_2$AgBi$_{0.5}$Fe$_{0.5}$Br$_6$.
}
\end{figure}

Crucially, this optimized growth protocol enabled a significant finding. In contrast to a previous report \cite{key13} which found that Fe doping introduces more disorder without altering the band gap, our high-quality Fe-CABB crystals exhibit a narrowed energy gap and reduced disorder below the phase transition, based on SC-XRD and PL data (shown in the following sections). We attribute this distinct behavior to the superior crystallinity achieved by our modified growth conditions.

\subsection{Temperature-Dependent Single-Crystal X-ray Diffraction}
\subsubsection{Structural Phase Transition from Cubic to Tetragonal}
To probe the crystal order in ${\mathrm{Cs_2AgBiBr_6}}$ and its Fe-doped variant ${\mathrm{Cs_2AgBi_{0.5}Fe_{0.5}Br_6}}$, temperature-dependent single-crystal X-ray diffraction (SC-XRD) was performed from 100~K to 225~K. The temperature evolution of the c-axis lattice parameter, determined using the Crystals refinement software, is shown in Figure 2. Consistent with previous reports \cite{key23,key25,key26}, a structural phase transition from cubic to tetragonal is observed at a transition temperature ($T_s$) of approximately 125~K, identified by a minimum in the \textit{c}-axis parameter.

\begin{figure}[ht]
    \centering
        \centering
        \includegraphics[width=0.6\linewidth]{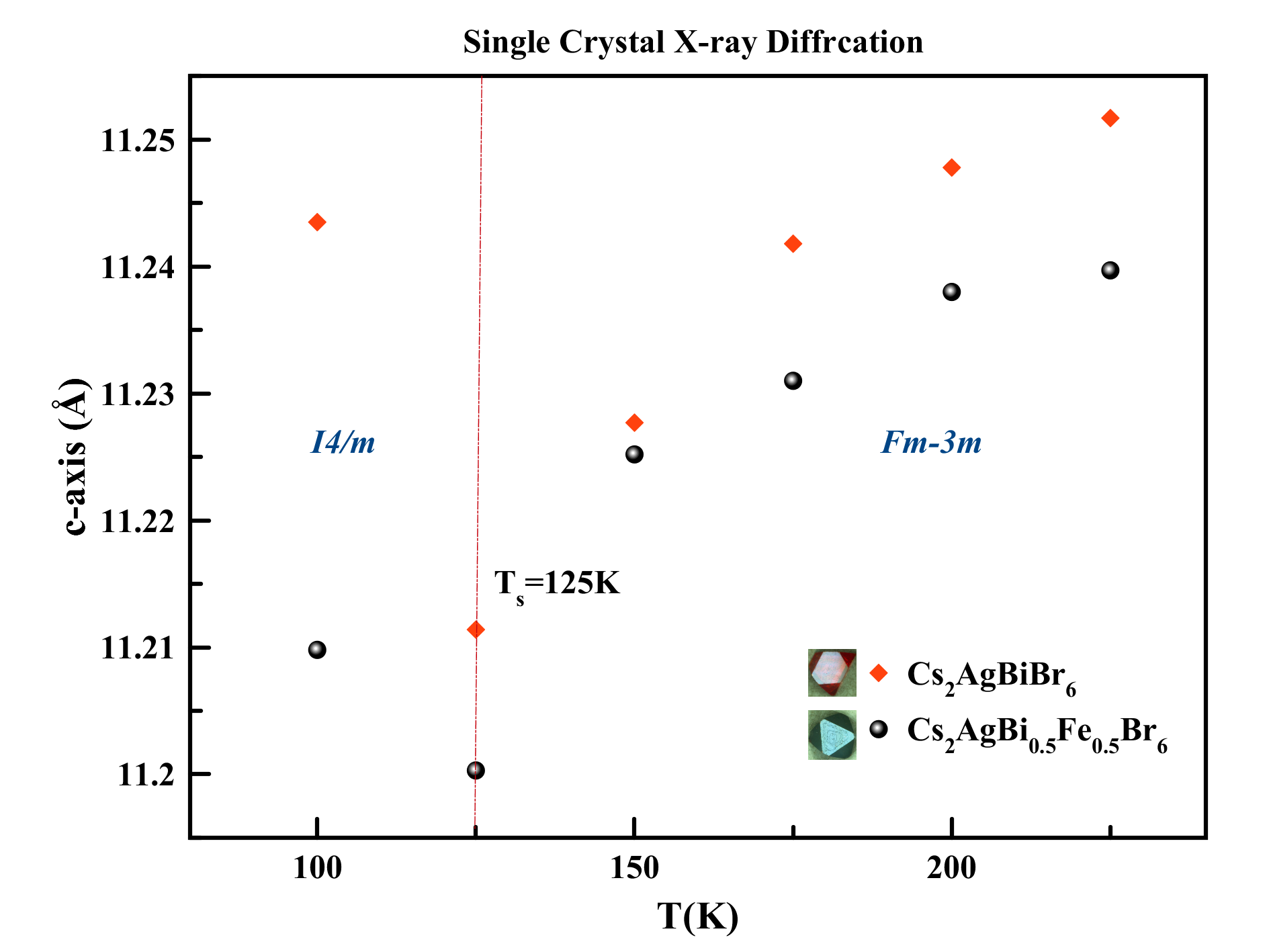}
    \caption{Temperature dependence of the c-axis lattice parameter obtained from single-crystal X-ray diffraction. The structural phase transition from cubic (FCC) to tetragonal ($I4/m$) is indicated at 125 K. Data for $\mathrm{Cs_2AgBiBr_6}$ and $\mathrm{Cs_2AgBi_{0.5}Fe_{0.5}Br_6}$ are represented by orange diamonds and black solid balls, respectively.
}
\end{figure}

Above $T_s$, ${\mathrm{Cs_2AgBiBr_6}}$ adopts a face-centered cubic crystal structure with the space group \textit{Fm}$\bar{3}$\textit{m} (No.~225), in Supplementary Figure S1a. Doping was achieved by substituting 50\% of the ${\mathrm{Bi^{3+}}}$ sites with ${\mathrm{Fe^{3+}}}$, resulting in the composition ${\mathrm{Cs_2AgBi_{0.5}Fe_{0.5}Br_6}}$, whose schematic structure is shown in Figure S1b. Upon cooling below $T_s$, both the parent and Fe-doped compounds undergo a structural phase transition to a tetragonal phase with the space group \textit{I4/m} (No.~87), depicted in Figures~S1c and~S1d, respectively. The following sections will provide a detailed comparison of the crystal structures for each phase.

\subsubsection{Mitigation of Strain and Structural Modification by Fe Doping}
The temperature dependence of the \textit{c}-axis lattice parameter, shown in Figure~2, reveals two key findings. First, both compounds exhibit positive thermal expansion, with the \textit{c}-axis increasing from $T_s$ to 225~K. This expansion is reflected in the metal-halide bond lengths, such as the  Bi/Fe--Br1 and Bi--Br1 bonds, which show a corresponding increase (e.g., from 2.8189~$\mathrm{\AA}$ to 2.8263~$\mathrm{\AA}$ in the parent compound). Second, the \textit{c}-axis of the Fe-doped crystal ($\mathrm{Cs_2AgBi_{0.5}Fe_{0.5}Br_6}$) is consistently smaller than that of $\mathrm{Cs_2AgBiBr_6}$ at all temperatures. This systematic lattice contraction is direct evidence of successful doping, arising from the substitution of larger $\mathrm{Bi^{3+}}$ ions (e.g., 1.03~$\mathrm{\AA}$) with smaller $\mathrm{Fe^{3+}}$ ions (e.g., 0.645~$\mathrm{\AA}$ for high-spin). The concomitant shortening of the average Bi/Fe--Br bond distance in $\mathrm{Cs_2AgBi_{0.5}Fe_{0.5}Br_6}$ provides further confirmation of Fe incorporation into the lattice. Since the cubic phase has been well studied from previous reports \cite{key23,key25}, we will focus on the new observations from the \textit{I4/m} phase as below.

To gain deeper insight into the crystal properties, a detailed structural analysis of the tetragonal phase at $100\,\mathrm{K}$ was performed using the Jana2006 software package; the results are summarized in Table 1. The phase transition from cubic ($a = b = c$) to tetragonal ($a = b < c$) symmetry induces a tetragonal strain within the \textit{ab}-plane.  This strain is quantified by $\sigma = 2(a-c)/(a+c)$, where a and c are the lattice parameters (Table 1). This parameter provides a normalized measure of deviation from cubic symmetry (where  $\sigma$ = 0). In our samples, the observed c-axis elongation with concomitant a-axis contraction yields negative values of  $\sigma$. The calculated values of $\sigma$ are $-0.3464$ for $\mathrm{Cs_2AgBiBr_6}$ and $-0.3416$ for the $\mathrm{Cs_2AgBi_{0.5}Fe_{0.5}Br_6}$ compound. The negative sign confirms a compressive strain along the \textit{a}/\textit{b}-axes. The larger absolute value of $|\sigma|$ in the parent compound indicates a stronger tetragonal deformation compared to the Fe-doped sample, suggesting that Fe doping helps maintain lattice order in the \textit{I4/m} phase and may suppress the formation of antisite defects (e.g., $\mathrm{Ag_{Bi}}$ or $\mathrm{Bi_{Ag}}$) commonly observed in $\mathrm{Cs_2AgBiBr_6}$.

This interpretation is supported by an analysis of the metal-halide bond distances (Table 1). In the pristine $\mathrm{Cs_2AgBiBr_6}$ structure, the relationship $d(\mathrm{Bi-Br1}) < d(\mathrm{Bi-Br2})$ is consistent with the tetragonal distortion where the lattice parameters satisfy $a, b < c$ (with Br1 at $\sim(0.25,0.77,0)$ and Br2 at $\sim(0,0,0.25)$). In contrast, the Fe-doped crystal exhibits a reversed inequality, $d'(\mathrm{Bi/Fe-Br1}) > d'(\mathrm{Bi/Fe-Br2})$. This distinct bonding environment is explained by Fe preferentially occupying Bi sites along the c-axis, leading to a more pronounced reduction in the $d'(\mathrm{Bi/Fe-Br2})$ bond length. This is consistent with the observed lattice parameter changes, where doping expands the a-axis and contracts the c-axis relative to the pristine crystal. The equivalence of the Ag–Br bond lengths between the two samples confirms that Fe substitution occurs exclusively at the Bi site, thereby suppressing the formation of Ag–Bi anti-site defects commonly observed in $\mathrm{Cs_2AgBiBr_6}$. The impact of this reduced defect density on the electronic properties is further evidenced by the narrower spectral linewidth (FWHM) observed in the photoluminescence (PL) data.

\begin{table}
\caption{\label{tab:table1}%
Lattice parameters for Cs$_2$AgBiBr$_6$ and Cs$_2$AgBi$_{0.5}$Fe$_{0.5}$Br$_6$ at 100 K in the tetragonal phase (space group $I4/m$).}

\begin{ruledtabular}
\begin{tabular}{lllll}
\textrm{sample}&
\textrm{a(=b)}&
\textrm{c}&
\textrm{Bi-Br$_1$}&
\textrm{Bi-Br$_2$}\\
\colrule
Cs$_2$AgBiBr$_6$ & 7.923(2) & 11.243(5) & 2.813(5) & 2.823(5) \\
Cs$_2$AgBi$_{0.5}$Fe$_{0.5}$Br$_6$ & 7.938(6) & 11.209(8) & 2.821(0) & 2.812(6)\\
\end{tabular}
\end{ruledtabular}
\end{table}
\begin{table}
\begin{ruledtabular}
\begin{tabular}{lllll}
\textrm{Ag-Br$_1$}&
\textrm{Ag-Br$_2$}&
\textrm{$\sigma$}&
\textrm{$\sigma_z$$/$$\sigma_{xy}$(Bi-Br)}&
\textrm{$\sigma_z$$/$$\sigma_{xy}$(Ag-Br)}\\
\colrule
2.790(5) & 2.798(5) & -0.3464 & 1.004 & 1.003\\
2.793(4) & 2.793(6) & -0.3416 & 1.003 & 1\\
\end{tabular}
\end{ruledtabular}
\end{table}

At the $I4/m$ phase, the enhanced structural order observed in $\mathrm{Cs_2AgBi_{0.5}Fe_{0.5}Br_6}$ is likely a direct result of the optimized crystal growth conditions, which provided a more favorable thermodynamic landscape for nucleation and growth. While the hydrothermal method is commonly used for synthesizing $\mathrm{Cs_2AgBiBr_6}$ single crystals \cite{key4,key13,key14}, our approach introduced key modifications: a pre-melting step at $180\,^{\circ}\mathrm{C}$ for one hour, annealing for 24 hours, and a significantly reduced cooling rate of $0.08\,^{\circ}\mathrm{C}/\mathrm{min}$ to promote larger crystal sizes (Figure 1). The pre-melting step is hypothesized to achieve a more homogeneous distribution of ions in the precursor solution. Subsequently, the slow cooling rate provides the necessary kinetic control, favoring the orderly substitution of $\mathrm{Bi^{3+}}$ by $\mathrm{Fe^{3+}}$ and their preferential stacking along the \textit{c}-axis. This hypothesis is consistent with reports on other systems, where similar thermodynamic control in hydrothermal synthesis has been shown to influence crystal size and minimize defect formation by lowering the overall free energy of the system \cite{key31}.

Thus, high-concentration Fe-doped $\mathrm{Cs_2AgBiBr_6}$ single crystals were synthesized via a modified crystal growth method. Single-crystal X-ray diffraction confirms a phase transition from a cubic to a tetragonal ($I4/m$) structure in both the pristine and doped compounds. The introduction of Fe in the cubic phase leads to a uniform lattice contraction; this structural modification is the direct cause of the observed band gap narrowing, an effect will be elucidated through photoluminescence (PL) analysis. Within the tetragonal phase, Fe doping effectively suppresses the strain disorder inherent to the pristine sample. A detailed analysis of metal-halide bond lengths demonstrates that Fe incorporates exclusively onto the Bi site, with a preferential alignment along the c-axis. This site-selective substitution alleviates local lattice strain and reinforces the host framework's structural integrity, as evidenced by the suppression of deleterious disorder between 90 and 125 K in PL measurements. The resulting robust and stable crystal structure is particularly advantageous for fabricating highly stable optoelectronic capable of operating at low temperatures.

\subsection{Photoluminescent Properties}

\subsubsection{Steady-State PL Spectra of Pristine and Doped Crystals}
The temperature-dependent photoluminescence (PL) spectra of pristine $\mathrm{Cs_2AgBiBr_6}$ and Fe-doped $\mathrm{Cs_2AgBi_{0.5}Fe_{0.5}Br_6}$ single crystals were measured from 300~K down to 80~K (Figure~3). A detailed analysis of the integrated PL intensity, peak position, and full width at half maximum (FWHM)---obtained from Gaussian fitting---is presented in Figure~4.

The evolution of all three PL parameters reveals a distinct anomaly centered at approximately 125~K for both samples. This temperature coincides with the structural phase transition (from cubic to tetragonal) temperature, $T_s$, which is well-established in literature \cite{key25, key26} and has been confirmed by our own SC-XRD measurements. Upon cooling from 300~K to T$_s$, the integrated PL intensity increases, reaching a maximum at T$_s$=125 K, then decreases upon further cooling (Figure 4a). Concurrently, the PL peak energy and spectral linewidth (FWHM) both exhibit a characteristic V-shaped temperature dependence (Figure 4b, c), with the transition region extending from the onset near 150~K to the completion near 90~K. The inflection point in this V-shaped dependence at ~125 K corresponds to the structural phase transition temperature, T$_s$ \cite{key25, key26}.

\begin{figure}
        \centering
        \includegraphics[width=1\linewidth]{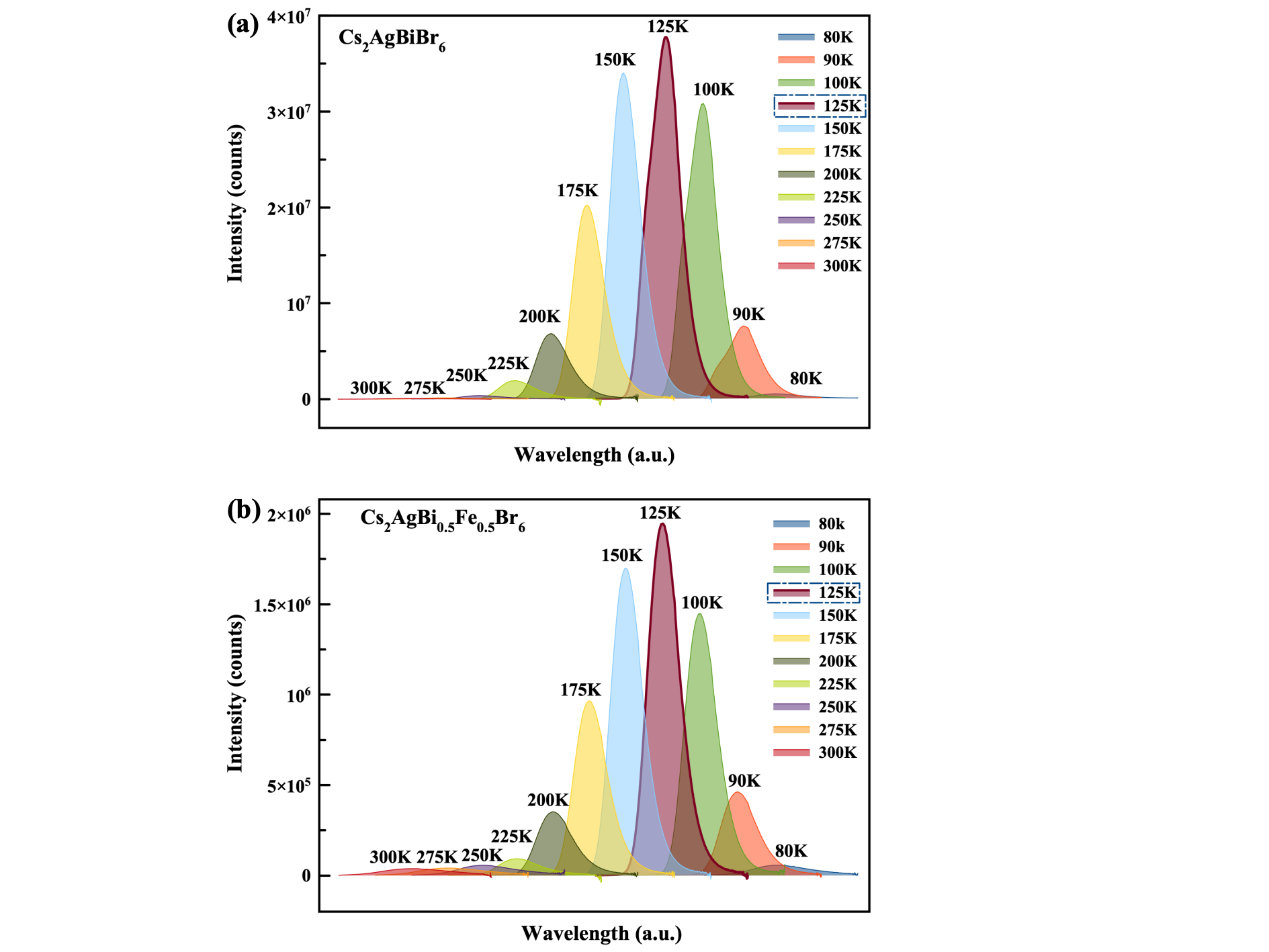}
    \caption{Temperature-dependent steady-state photoluminescence (PL) spectra of (a) pristine $\mathrm{Cs_{2}AgBiBr_{6}}$ and (b) $\mathrm{Cs_{2}AgBi_{0.5}Fe_{0.5}Br_{6}}$ single crystals, measured from 80 K to 300 K.
}
\end{figure}

\begin{figure}
        \centering
        \includegraphics[width=1\linewidth]{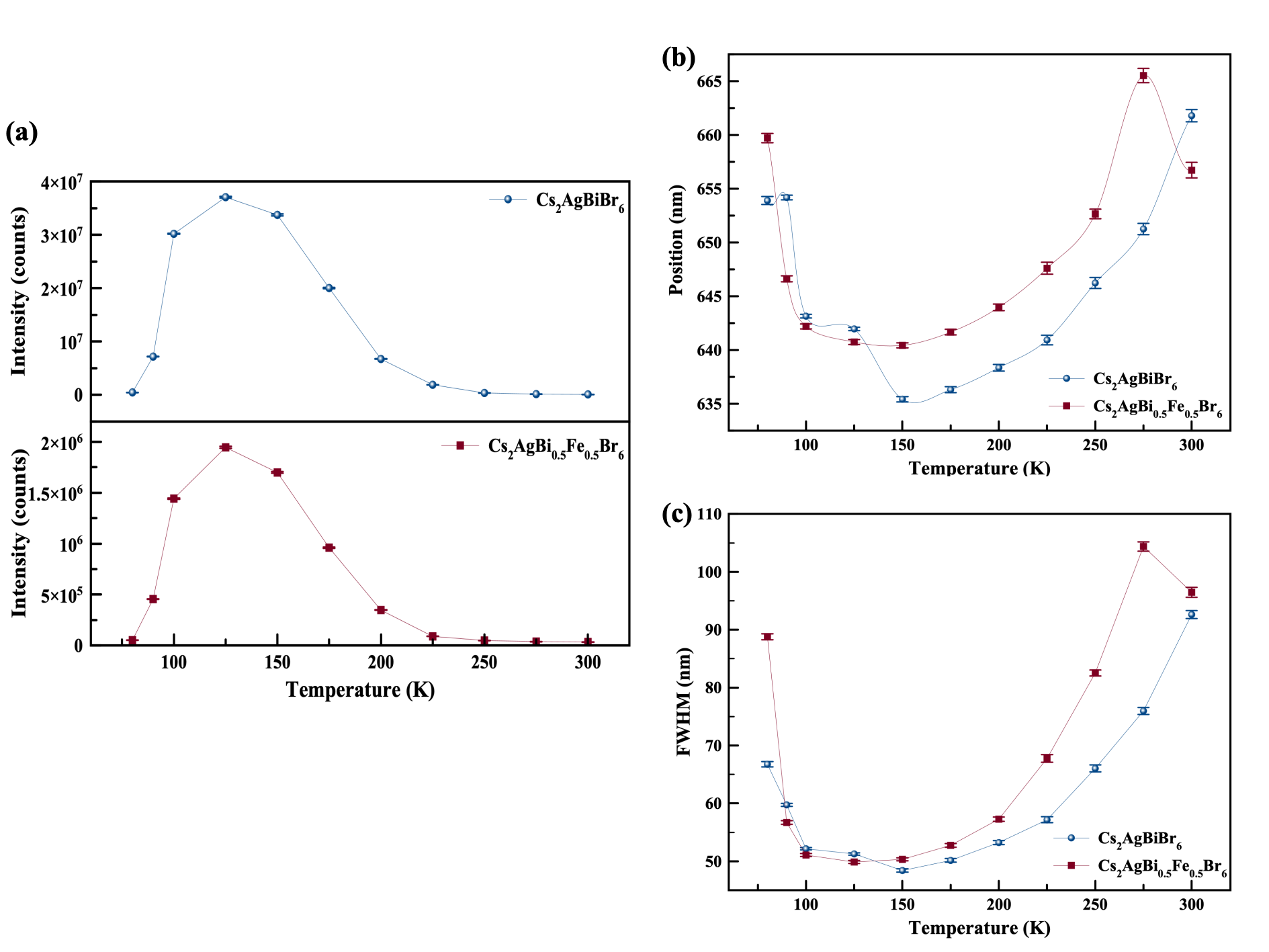}
    \caption{Temperature dependence of the PL spectra analysis for (a) integrated peak intensity, (b) peak position, and (c) full width at half maximum (FWHM). Data for pristine $\mathrm{Cs_{2}AgBiBr_{6}}$ and $\mathrm{Cs_{2}AgBi_{0.5}Fe_{0.5}Br_{6}}$ are represented by blue solid balls and dark red squares, respectively.
}
\end{figure}

A direct comparison between the pristine and Fe-doped samples highlights several key doping effects. First, the integrated PL intensity of the pristine crystal is approximately an order of magnitude greater across the entire temperature range, which we attribute to Fe doping introducing non-radiative recombination centers. Second, the PL peak position (Figure 4b) shows a significant trend: between 300 K and  $T_s$, the Fe-doped sample emits at a consistently red-shift (a shift to lower energy), indicating a successful narrowing of the band gap. Critically, this trend reverses below  $T_s$, where the Fe-doped sample emits at a similar or slightly higher energy than the pristine sample. Finally, the spectral linewidth (FWHM) (Figure 4c) exhibits a complementary crossover: the Fe-doped sample has a larger FWHM above  $T_s$ but a narrower one below it compared to the pristine sample, implying reduced strain disorder after the transition. Notably, the 300 K data stand out as a key exception. The initial differences in peak position and FWHM originate from local structural instabilities induced by Fe doping in the high-temperature phase. Our higher-temperature Raman studies reveal that these instabilities vanish upon cooling, allowing the underlying crossover trends to emerge across $T_s$. In contrast, the anomaly in the 80 K data for both peak position and FWHM suggests a possible crossover in the dominant lattice dynamics at low temperature, potentially related to host-lattice stabilization following the structural phase transition. The origin of this single-point anomaly requires further investigation.

\subsubsection{Effect of Fe Doping on PL Emission}
We interpret these results as evidence of two distinct roles of Fe doping, corresponding to the two structural phases. In the high-temperature cubic phase, Fe acts primarily as an electronic dopant, introducing band gap states that narrow the effective band gap and increase carrier scattering (evidenced by the red-shift and larger FWHM). This is consistent with the third harmonic generation (THG) measurements and density functional theory (DFT) calculations reported elsewhere, which show that Fe-3d states form an intermediate band within the pristine band gap, located near the Fermi level, thereby narrowing the effective band gap \cite{key13}. In the low-temperature tetragonal (\textit{I4/m}) phase, the influence of the structural transition dominates. The convergence and crossover of the peak energies suggest that the structural distortion—particularly the expansion of the \emph{c}-axis—has a more pronounced effect on the band gap of the pristine crystal. This is corroborated by our earlier SC-XRD data, which indicates that Fe doping suppresses strain distortion, thereby enhancing structural stability in the \textit{I4/m} phase. 

Consequently, the smaller FWHM in the Fe-doped sample below $T_s$ indicates higher crystalline quality in this phase, likely due to suppressed dynamic disorder or electron-phonon coupling. The ability of Fe doping to tune the band gap in one phase and enhance structural stability in the other, while simultaneously influencing carrier scattering pathways, highlights its potential for controlling optoelectronic properties in device applications.

\subsubsection{Contrasting Low-Temperature PL: Splitting and its Suppression by Fe}

Our discussion of the PL trends in the above text is based on a single-Gaussian analysis, which provides a robust and consistent basis for comparing the average electronic properties of the pristine and Fe-doped samples. However, the observed asymmetry in the emission spectrum—attributed to self-trapped excitonic species \cite{key28,key29}—suggests a more complex underlying mechanism. To gain deeper insight into the underlying emission mechanisms, we performed a detailed multi-peak lineshape analysis. As detailed in the supplementary Figure 2S, a two-peak model successfully captures the phase transition at  $T_s$ $\approx 125$~K for both samples. A key difference emerges below $T_s$, however: the pristine sample requires a three-peak model, while the Fe-doped sample remains well-described by only two components, see Figure 5.

\begin{figure}
    \centering
    \begin{minipage}{0.45\textwidth}
        \centering
        \includegraphics[width=\textwidth]{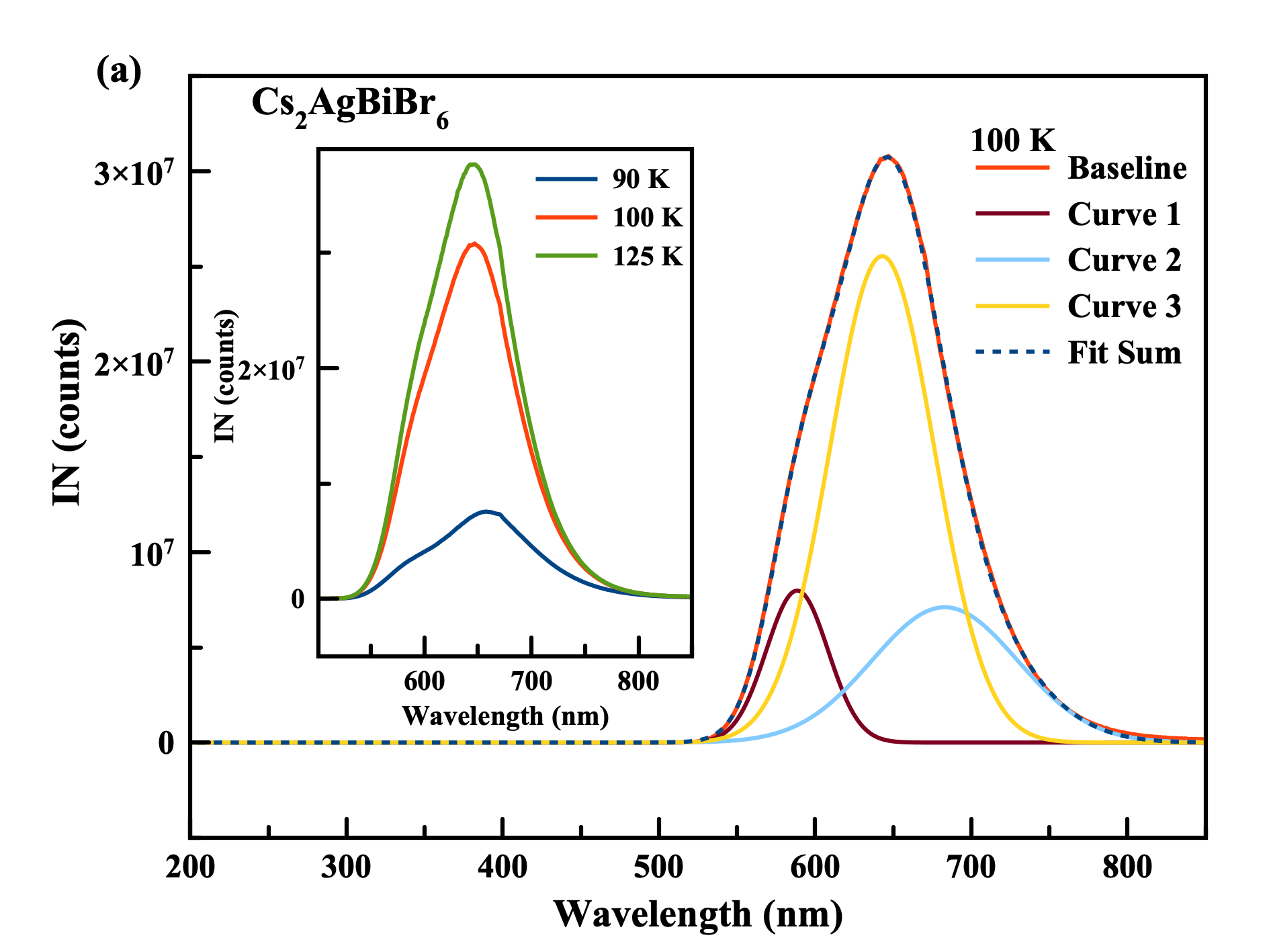}
        \label{fig:graph1}
    \end{minipage}
    \hfill
    \begin{minipage}{0.45\textwidth}
        \centering
        \includegraphics[width=\textwidth]{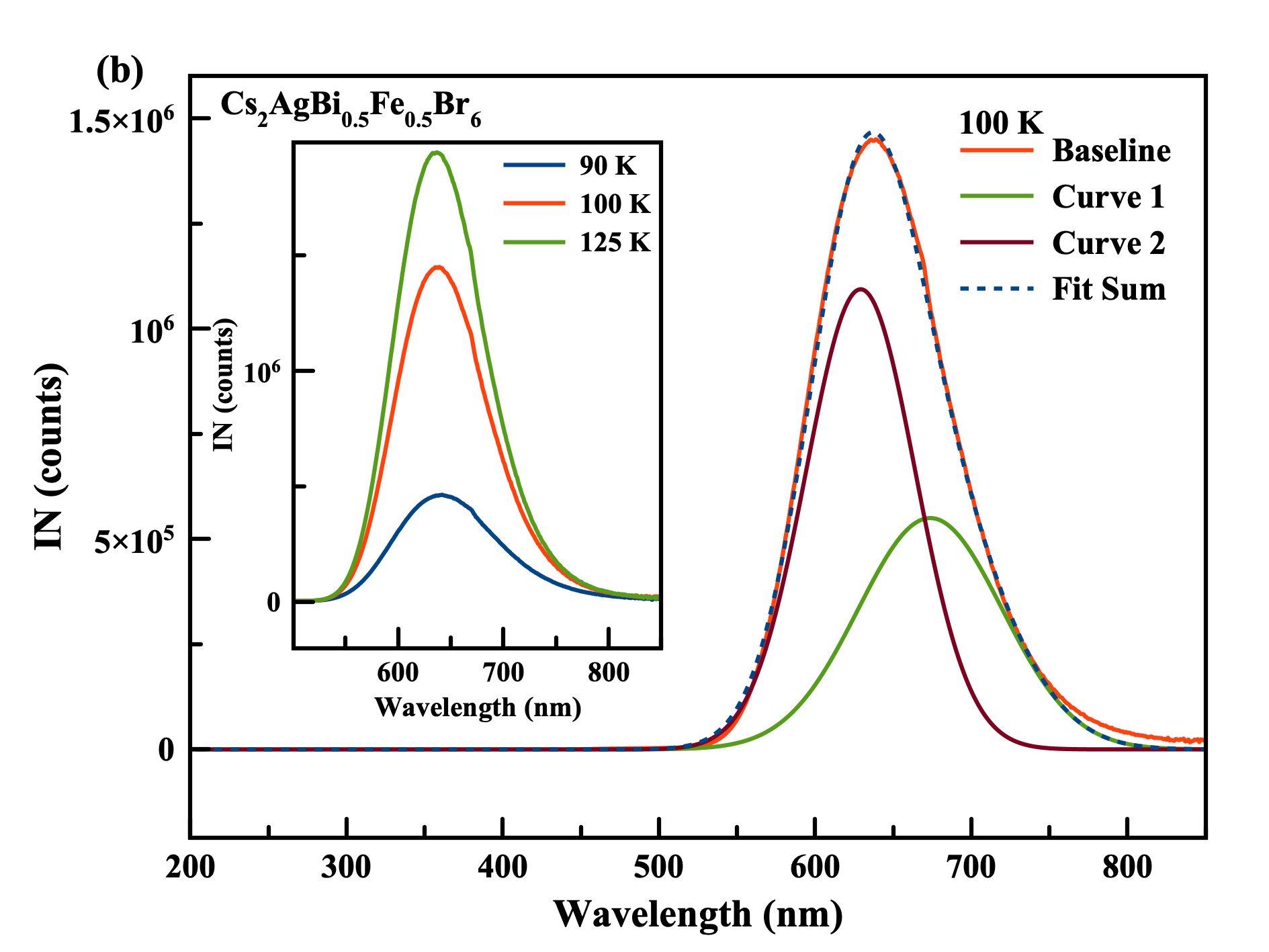}
        \label{fig:graph2}
    \end{minipage}
    \caption{Multi-peak Gaussian fitting of the PL spectra at 100 K for (a) $\mathrm{Cs_2AgBiBr_6}$, requiring three components, and (b) $\mathrm{Cs_2AgBi_{0.5}Fe_{0.5}Br_6}$, requiring two components. The insets show the corresponding spectra from 90 K to 125 K. In both main panels, the orange line represents the baseline and the blue line is the sum of the fitted components.
}
\end{figure}

The striking transient peak splitting was observed in the pristine sample immediately below the phase transition (90–125 K), shown in Figure 5a, which vanished upon further cooling to 80 K. This feature was entirely absent in the Fe-doped sample (Figure 5b). We attribute this transient splitting to stronger lattice vibrations and distortions caused by a more substantial lattice parameter change in the pristine material near its phase transition. Our earlier strain analysis at 100 K directly has proved this, confirming that the Fe-doped sample undergoes a smaller change in strain (Table 1).

These collective findings demonstrate that Fe doping plays a crucial role in stabilizing the host crystal structure, particularly across the phase transition. This enhanced structural robustness is highly promising for applications in low-temperature optoelectronics. The incorporation of Fe into $\mathrm{Cs_{2}AgBiBr_{6}}$ thus presents a viable and effective strategy for engineering more stable functional materials capable of operating reliably under cryogenic conditions.

\subsubsection{Time-Resolved Photoluminescence}

Time-resolved photoluminescence (TRPL) decays (90–300 K) for both pristine and Fe-doped $\mathrm{Cs_2AgBiBr_6}$ single crystals were best fitted with a tri-exponential function. This analysis yielded three distinct lifetime components—$\tau_1$, $\tau_2$, and $\tau_3$ —with the results displayed in Figure 6 and representative decay curves provided in the supplementary materials Figure S3.

\begin{figure}[t]
    \centering
    \begin{minipage}{0.45\textwidth}
        \centering
        \includegraphics[width=\textwidth]{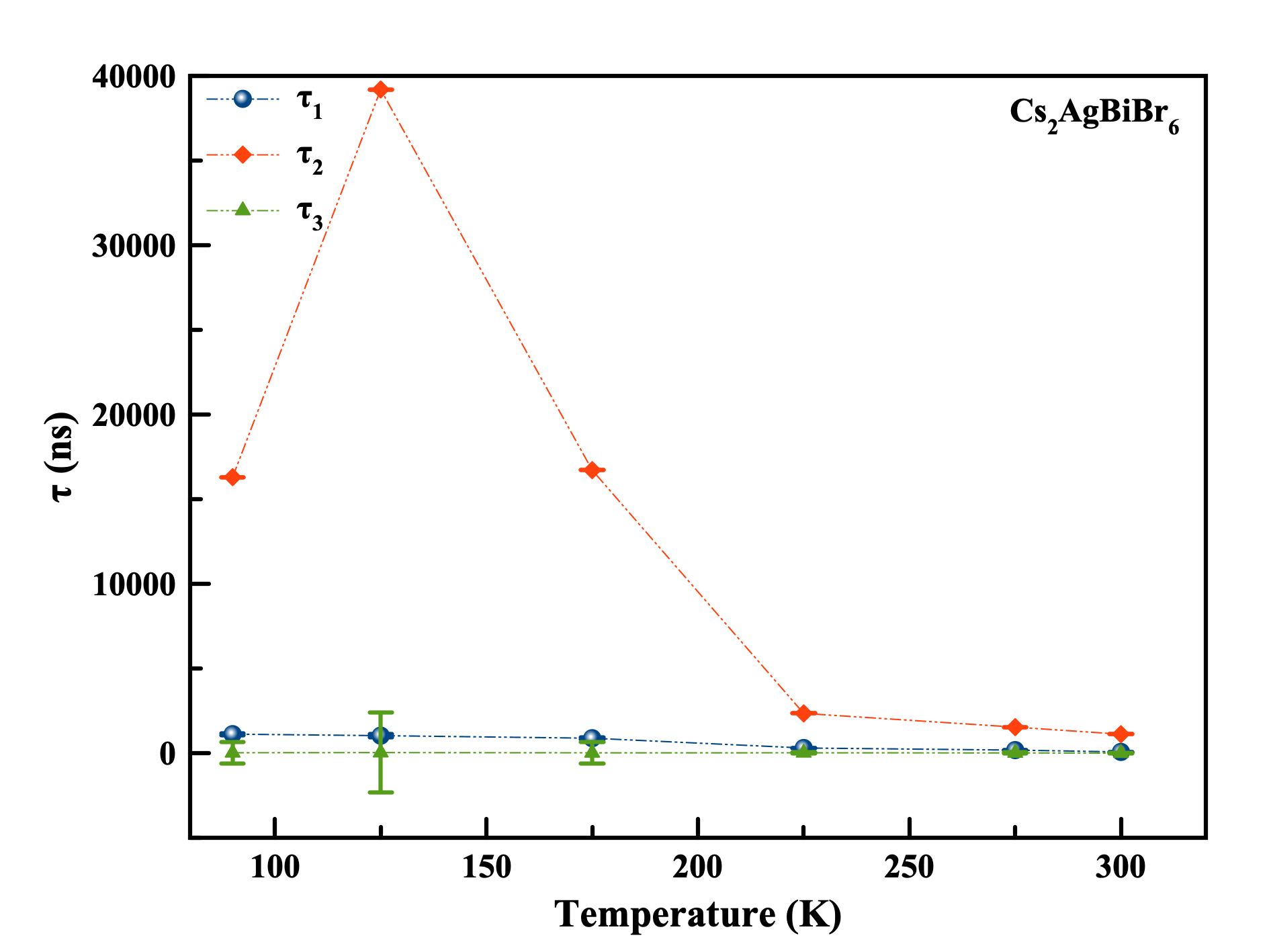}
        \label{fig:graph1}
    \end{minipage}
    \hfill
    \begin{minipage}{0.45\textwidth}
        \centering
        \includegraphics[width=\textwidth]{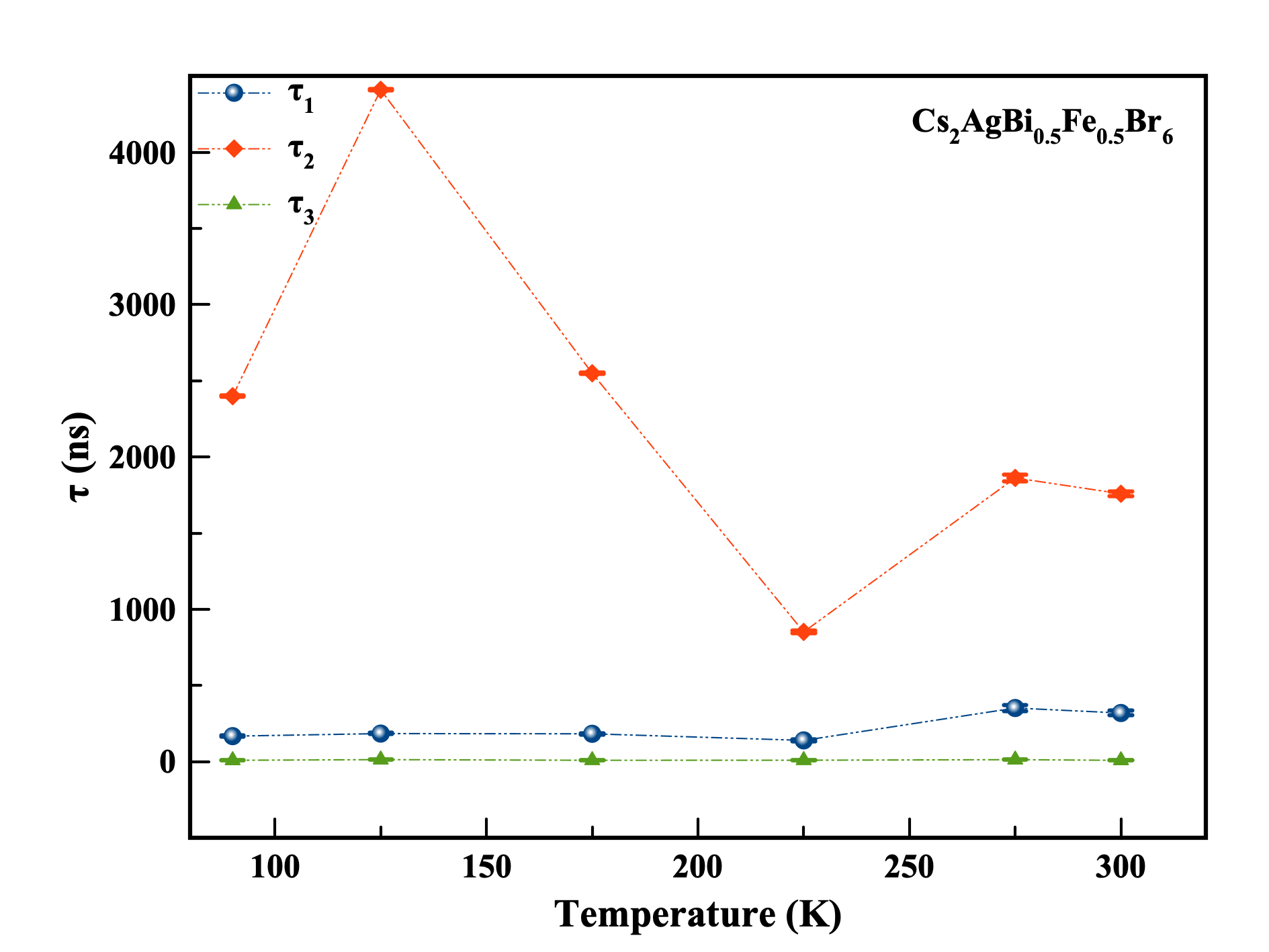}
        \label{fig:graph2}
    \end{minipage}
    \caption{Temperature-dependent lifetimes ($\tau_1$, $\tau_2$, and $\tau_3$) extracted from tri-exponential fits to the time-resolved photoluminescence (TRPL) decays. The data are represented by blue solid balls ($\tau_1$), orange diamonds ($\tau_2$), and green triangles ($\tau_3$).
}
\end{figure}

A pronounced anomaly at approximately 125~K was observed exclusively in the temperature dependence of the long-lived lifetime component, $\tau_{2}$, for both samples. This anomaly coincides with the structural phase transition, as identified by the SC-XRD and the PL intensity measurements. In contrast, $\tau_{1}$ and $\tau_{3}$ showed no significant temperature dependence. The strong correlation between the integrated PL intensity and $\tau_{2}$ at this transition identifies it as the radiative-dominated lifetime component. This assignment is consistent with previous reports that attribute the long-lived emission in this material class to self-trapped excitons (STEs) \cite{key5,key27,key30}, confirming that $\tau_{2}$ corresponds to intrinsic STE recombination in the bulk crystal. Consequently, the temperature-independent $\tau_{1}$ and $\tau_{3}$ components are attributed to trap states or defect-related emissions \cite{key17}.

While a direct comparison confirms that Fe doping introduces efficient non-radiative pathways—evidenced by a tenfold reduction in the $\tau_{2}$ lifetime compared to the pristine sample—the key finding of this work is the superior quality of both samples synthesized by our method. Critically, the $\tau_{2}$ lifetime of our Fe-doped sample remains comparable to, or even exceed, the values reported in the literature for this kind of compounds \cite{key5,key17,key12}. This demonstrates that the high crystal quality we achieve can mitigate the typical detrimental effects of doping to produce a viable material. Furthermore, the well-defined maximum in the $\tau_{2}$ lifetime at 125 K provides a clear road-map for optimal device operation. 

Though the observed non-radiative recombination suggests Fe-doped CABB is unsuitable for high-efficiency photovoltaics, its enhanced structural stability during phase transitions, reduced bandgap in the cubic phase, and low-cost, non-toxic nature make it a compelling material for applications where temperature resilience is critical. Promising directions include robust X-ray detectors and spin-based (spintronic) devices.

In summary, the PL analysis demonstrates that Fe doping in $\mathrm{Cs_2AgBiBr_6}$:
\begin{itemize}
\item[\textbf{a.}] Preserves the host crystal structure and its phase transition, thereby enhancing the material's structural robustness.
\item[\textbf{b.}] Effectively reduces the band gap in the high-temperature phase, directly addressing a key limitation for the application of lead-free double perovskites in optoelectronics and spintronics.
\item[\textbf{c.}] Helps maintain crystal stability below and across the phase transition, as evidenced by the suppression of transient spectral splitting and the modified carrier-lattice coupling observed in the linewidth analysis.
\end{itemize}

\subsection{Probing the Local Lattice Environment: Near-Room-Temperature Raman Spectroscopy }
\begin{figure}[htbp]
        \centering
        \includegraphics[width=0.9\linewidth]{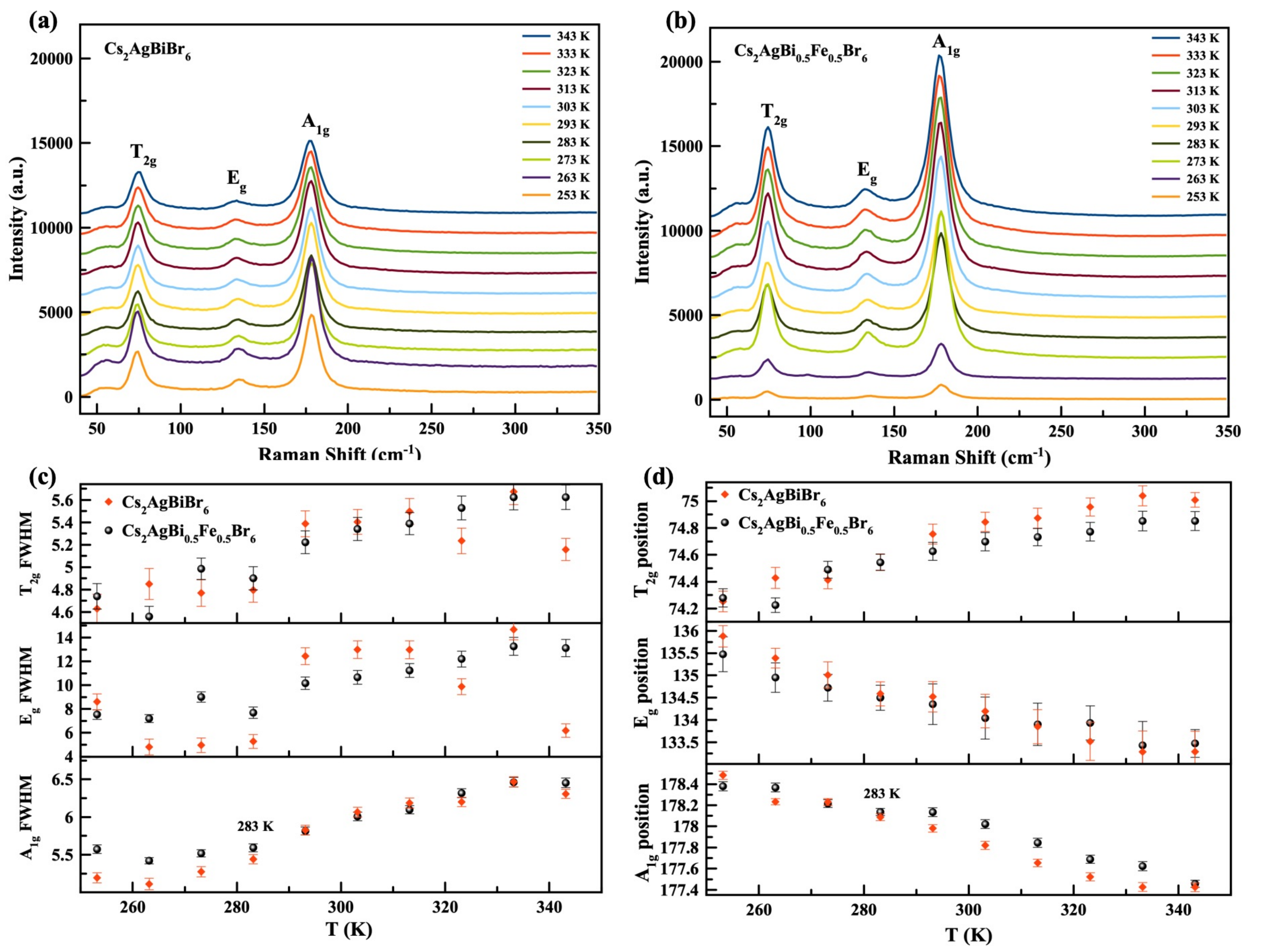}
    \caption{(a, b) Temperature-dependent Raman spectra near room temperature for (a) $\mathrm{Cs_2AgBiBr_6}$  and (b) $\mathrm{Cs_2AgBi_{0.5}Fe_{0.5}Br_6}$. (c, d) Analysis of the Raman spectra showing the temperature dependence of (c) the FWHM and (d) the peak position of the dominant mode, obtained from Gaussian fitting. Data for $\mathrm{Cs_2AgBiBr_6}$ and $\mathrm{Cs_2AgBi_{0.5}Fe_{0.5}Br_6}$ are represented by orange diamonds and black solid balls, respectively.
}
\end{figure}

To elucidate the anomalous 300 K photoluminescence (PL), temperature-dependent Raman spectroscopy (Figure 7) was conducted on $\mathrm{Cs_2AgBiBr_6}$ and $\mathrm{Cs_2AgBi_{0.5}Fe_{0.5}Br_6}$ single crystals over a range of $253$~K to $343$~K. Both compositions exhibit the standard phonon modes (Figures 7a,b) of the $\mathrm{Cs_2AgBiBr_6}$ structure (T$_{2g}$, E$_g$, A$_{1g}$) \cite{key15,key32,key33}, confirming that Fe doping does not alter the fundamental crystal symmetry.

While the phonon mode linewidths (FWHM) and positions (Figures 7b,c) for the two samples are largely similar, their temperature-dependent trends reveal a key difference: the values for the Fe-doped sample show a consistent shift relative to the pristine crystal, which becomes apparent around $283$~K—precisely in the temperature range of the PL anomaly at 300 K (see Figures 4b,c). This divergence in phonon behavior provides direct evidence that Fe doping introduces dynamic lattice instabilities near room temperature. These instabilities are suppressed at lower temperatures, where Fe doping acts to shrink the lattice and tune the energy gap. Therefore, the Raman data confirm that while Fe doping preserves the host structure, it induces a temperature-sensitive local distortion that stabilizes upon cooling.

\section{Conclusion}

This study establishes Fe doping as a powerful, dual-functional strategy for engineering lead-free double perovskite $\mathrm{Cs_2AgBiBr_6}$. We demonstrate that our Fe-doped crystals (nominal $\mathrm{Cs_2AgBi_{0.5}Fe_{0.5}Br_6}$) simultaneously suppress the intrinsic structural instability below the 125 K phase transition and reduce the band gap above it. Critically, Fe doping achieves this enhanced low-temperature stability without introducing a new structural distortion, but by suppressing the deleterious strain and disorder that characterize the phase transition in the pristine lattice. Our work, enabled by high-quality single-crystal synthesis, transforms $\mathrm{Cs_2AgBiBr_6}$ into a robust, low-gap semiconductor, paving the way for its practical application in next-generation optoelectronics and spintronics.

\section*{Methods}
\textbf{Single Crystal Synthesis:}
Single crystals of $\mathrm{Cs_2AgBiBr_6}$ and $\mathrm{Cs_2AgBi_{0.5}Fe_{0.5}Br_6}$ were grown using a hydrothermal method identical to that reported in our previous work \cite{key15}, employing the precursor ratios detailed in Table 1 (entry for x = 0.0) of the associated supplementary materials. The synthesis of iron-doped $\mathrm{Cs_2AgBi_{0.5}Fe_{0.5}Br_6}$ followed a similar procedure and precursor ratio, with the modification that iron(III) bromide (FeBr$_3$) powder was introduced to the precursor mixture in a 1:1 molar ratio relative to BiBr$_3$. The detailed crystal growth protocol is schematically represented in the manuscript Figure 1.

\textbf{Single-Crystal X-ray Diffraction (SC-XRD):}
Single-crystal X-ray diffraction data for $\mathrm{Cs_2AgBiBr_6}$ and $\mathrm{Cs_2AgBi_{0.5}Fe_{0.5}Br_6}$ were collected on a Rigaku XtaLAB Synergy R DW system equipped with a HyPix diffractometer and using Mo K$\alpha$  radiation ($\lambda$ = 0.71073 $\mathrm{\AA}$). The crystal was mounted on a cryo-loop and kept at a steady temperature of 99 K during data collection using the instrument's cryostat system. To investigate temperature-dependent structural changes, data sets were also collected at 125, 150, 175, 200, and 225 K. The lowest achievable temperature on this instrument is 100 K. The data were integrated and scaled using the Crystals and Jana2006 software package. The structure was solved by intrinsic phasing (SHELXT).

\textbf{Photoluminescence (PL) Spectroscopy:}
Steady-state photoluminescence characterization was performed using an Edinburgh Instruments FLS1000 fluorescence spectrophotometer, equipped with a xenon lamp excitation source and a photomultiplier tube (PMT) detector. Temperature-dependent PL measurements were conducted from 80 K to 300 K using a nitrogen gas cryostat.

\textbf{Raman Spectroscopy:}
Micro-Raman spectroscopy was performed on single crystals of $\mathrm{Cs_2AgBiBr_6}$ and $\mathrm{Cs_2AgBi_{0.5}Fe_{0.5}Br_6}$ using a Horiba LabRAM HR Evol instrument with measurement along $[1,1,1]$ direction. Spectra were acquired using a 633 nm HeNe laser with a power of 17 mW at the sample, a 600 l/mm grating, and a 50× objective, resulting in a laser spot size of approximately 1.5 $\mu$m. The acquisition time for a single spectrum was 3 seconds. For temperature-dependent studies, a Linkam cold stage was used to control the sample temperature from 253 K to 343 K.
\section*{Funding}
The authors gratefully acknowledge the New Faculty Founding Support from Liaoning Technical University to Y.L.

\section*{Declaration of competing interest}
The authors declare that they have no known competing financial interests or personal relationships that could have appeared to influence the work reported in this paper. 
\section*{Data availability}
Data is available upon request. 

\begin{acknowledgments}
 The authors thank the staff at the Center for Micro-Nano Fabrication and Advanced Characterization and the Instrument Center of the Materials Science Department at Xiamen University for their scientific and technical assistance.
\end{acknowledgments}

\bibliography{apssamp}

\end{document}